\documentclass[twoside, a4paper, nofootinbib,11pt]{article}
\usepackage{amssymb}
\usepackage{IEEEtrantools}
\usepackage[mathscr]{eucal}
\usepackage[dvips]{graphicx}
\usepackage{amsmath}
\usepackage{amsthm}
\usepackage{pstricks}
\usepackage{caption}
\def \ni{\noindent}

\newcommand{\be}{\begin{equation}}
\newcommand{\ee}{\end{equation}}
\newcommand{\ben}{\begin{equation*}}
\newcommand{\een}{\end{equation*}}
\newcommand{\bes}{\begin{eqnarray}}
\newcommand{\ees}{\end{eqnarray}}
\newcommand{\besn}{\begin{IEEEeqnarray*}{rCl}}
\newcommand{\eesn}{\end{IEEEeqnarray*}}

\newcommand{\txt}{\textrm}

\newtheorem{theorem}{Theorem}
\newtheorem*{theorem*}{Theorem}

\newtheorem*{definition*}{Definition}
\newtheorem*{lemma*}{Lemma}
\newtheorem*{prop*}{Proposition}
\newtheorem*{corollary*}{Corollary}
\newtheorem{definition}{Definition}
\newtheorem{prop}{Proposition}

\title{On the Preservation of Quasilocality by the Integration-Out Transformation}
\author{T. Tlas}
\date{}

\begin{document}
\maketitle
\thispagestyle{empty}

\begin{abstract}
\ni We demonstrate that the integration-out step of the renormalization group transformation preserves the quasilocality of the effective action. This is shown in the case of a single, real, scalar field on a torus, but the proof holds more generally. The main result can be thought of as showing the flow invariance of the quasilocal subset under the flow generated by the Polchinski equation.
\end{abstract}

\section{Introduction}

One of the deepest ideas in physics is the renormalization group. This is the concept that one can study a complicated system by analyzing it one scale at a time.\footnote{A very nice review is \cite{bk}. Moreover, it contains another version of (a part of) the results presented here for different conditions on the effective actions.}  Intuitively, it is clear that for this procedure to be practical, one should deal with systems which are at least approximately local, see the discussion in \cite{costello}. In technical terms, the effective actions must admit some form of derivative expansion. In other words, when one implements the renormalization group, a standard requirement is that the effective actions considered are required to be ``quasilocal'' using the terminology of \cite{rosten}. Roughly speaking, (we will give precise definitions later) these are functionals of the form

\ben
S[ \phi] = \sum_{n=0}^\infty \int dp_1 \dots dp_n \, \delta(p_1 + \dots + p_n) \, G(p_1, \dots, p_n) \, \phi(p_1) \dots \phi(p_n).
\een 

The goal of this paper is to prove rigorously that the integration-out step of the renormalization group preserves quasilocality. We shall demonstrate this fact in a specific realization, notably for a theory of a single real scalar field. However, it will be clear that the proofs can be easily generalized to other set-ups. \newline

As will be apparent below, one way to view the main result of this work is as a fact in the theory of viability \cite{pavel, vrabie}. This is a vast subject in which a great amount of work has been expended. However, the greatest majority of the established results pertain to conditions one puts on the full flow corresponding to the evolution equation (typically, a form of tangency of the flow to the set is considered). There are only a few results \cite{barbu, daprato} which supply sufficient conditions on the generator, and even then, they require information on the behaviour of the generator away from the set which is to be invariant. Below, a viability fact is proven under the assumption that the linear part of the equation preserves the set, while the nonlinear part of the generator maps the set to itself (see the discussion at the end of the paper for more details).

\section{Gaussian Measures}

As mentioned above, we will deal with the theory of a real scalar field $\phi$. To make things rigorous, we need both an infrared and an ultraviolet cutoff. We shall implement the infrared cutoff by putting periodic boundary conditions, i.e. we will consider our space of fields to be a certain space of functions on the $d$-dimensional torus $\mathbb{T}^d = \underbrace{S^1 \times S^1 \times \dots \times S^1}_{d \txt{ times}}$. The ultraviolet cutoff will be implemented by requiring the space of fields to consist of very smooth elements. Here is the precise definition:

\begin{definition}
Let $\{a_n\}_{n \in \mathbb{Z}^d}$ be a `sequence' of complex numbers. Define the norm $||| \cdot |||$ of such a sequence by

\ben
||| a_n  ||| = \sum_{n \in \mathbb{Z}^d} e^{\sqrt{n_1^2 + \dots + n_d^2}} \, |a_n|^2.
\een

Then the set of all those complex valued sequences $\{a_n\}_{n \in \mathbb{Z}^d}$ for which $||| a_n |||< \infty$ is a real Hilbert space with this norm. The corresponding inner product will be denoted by $<< \cdot, \cdot >>$. Let $V$ stand for the subspace consisting of all those elements which satisfy $a_{-n} = \overline{a_n}$.
\end{definition}

As it will be useful later on, let us indicate an orthonormal basis for $V$. To this end, let $\mathcal{B} \subset \mathbb{Z}^d$ stand for the following set:

\ben
\mathcal{B} =  \bigcup_{i=1}^{d} \bigg (  \bigcup_{n_i \in \mathbb{N}, (n_{i+1}, \dots, n_d) \in \mathbb{Z}^{d-i} } (0, \dots, 0, n_i, \dots, n_d)    \bigg).
\een

Now, an orthonormal basis of $V$ consists of the following collection of sequences $\{e_0 \} \cup \bigcup_{b \in \mathcal{B}} (e^+_b \cup e^-_b)   $, where

\besn
 e_0 &   \equiv & \{ \delta_{n_1, 0} \times \dots \times \delta_{n_d,0} \}_{(n_1, \dots, n_d) \in \mathbb{Z}^d}  \\
 e^+_b & \equiv & \frac{e^{- \frac{|n|}{2}}}{\sqrt{2}} \{ \delta_{n_1, b_1} \times \dots \times \delta_{n_d, b_d}  +  \delta_{n_1, - b_1} \times \dots \times \delta_{n_d, - b_d}   \}_{(n_1, \dots, n_d) \in \mathbb{Z}^d} \\
 e^-_b & \equiv & \frac{ i e^{- \frac{|n|}{2}}}{\sqrt{2}} \{ \delta_{n_1, b_1} \times \dots \times \delta_{n_d, b_d}  -  \delta_{n_1, - b_1} \times \dots \times \delta_{n_d, - b_d}   \}_{(n_1, \dots, n_d) \in \mathbb{Z}^d}.
\eesn

If we now use the usual Fourier correspondence

\besn
\phi(x) & \longleftrightarrow & a_n = \int_{ \mathbb{T}^d} \phi(x) e^{ -  i n x} dx \\
\{a_n \}_{n \in \mathbb{Z}^d} & \longleftrightarrow & \phi(x)  =    \sum_{n \in \mathbb{Z}^d} a_n e^{ i n x} ,
\eesn

where $nx = n_1 x_1 + \dots + n_d x_d$, and the measure on $\mathbb{T}^d$ is normalized so that it has volume $\frac{1}{(2 \pi)^d}$, we obtain at once that there is a linear bijection between $V$ and a certain subspace of the smooth, real-valued functions $\phi$ on the torus. This Hilbert space of functions, topologized by its isomorphism with $V$, is the space of fields on which all our measures will be defined. \newline

Our goal is to give a rigorous meaning to the following heuristic expression for the path integral measure

\ben
\exp \Big  [  -\frac{1}{2} \int_{\mathbb{T}^d} (   (\nabla \phi)^2(x) + \phi^2(x)) dx \Big ] \mathcal{D} \phi.
\een

We would like the following formula to hold

\besn
\int \exp\Big  [ i \int_{\mathbb{T}^d} j(x) \phi(x) dx \Big ] \exp\Big [  -\frac{1}{2} \int_{\mathbb{T}^d} (    (\nabla \phi)^2(x) + \phi^2(x)) dx \Big ] \mathcal{D} \phi & \sim & \\
\exp \Big [ - \frac{1}{2} \int_{\mathbb{T}^d} j(x) ( \frac{1}{ - \nabla^2 + 1} j(x) ) dx  \Big ] =  \exp \Big [- \frac{1}{2} \sum_{n \in \mathbb{Z}^d} \frac{|j_n|^2}{n^2 + 1} \Big  ], & &
\eesn

where we've used the Fourier isomorphism for the last equality. It turns out that it is not possible to define a Gaussian measure on $V$ such that the integral of $\exp \big [i \int_{\mathbb{T}^d} j(x) \phi(x) dx \big ]$ is equal to $\exp \big [- \frac{1}{2} \sum_{n \in \mathbb{Z}^d} \frac{|j_n|^2}{n^2 + 1} \big  ] $. However, by imposing ultraviolet cutoffs, we \textit{can} define regularized versions of the measure we want. In fact we have the following

\begin{prop}
Let $\Lambda, \Lambda'$ be positive numbers with $\Lambda > \Lambda'$. Then there exist Borel measures $\mu_\Lambda$ and $\mu_{\Lambda, \Lambda'}$ on $V$ such that for every $j \in V$ we have
\bes
\label{eqns:characteristic}
\int_V  e^{  i \int_{\mathbb{T}^d} j(x) \phi(x) dx } d \mu_\Lambda[\phi] & = & \int_V e^{i \sum_{n \in \mathbb{Z}^d} \overline{j_n} a_n} d \mu_\Lambda[\phi]  =   e^{- \frac{1}{2}K_\Lambda \big (\{j_n\}, \{ j_n \} \big )} \\
\int_V  e^{  i \int_{\mathbb{T}^d} j(x) \phi(x) dx } d \mu_{\Lambda, \Lambda'}[\phi] & = & \int_V  e^{  i \int_{\mathbb{T}^d} j(x) \phi(x) dx } d \mu_{\Lambda, \Lambda'}[\phi]  =  e^{- \frac{1}{2}K_{\Lambda, \Lambda'} \big (\{j_n\}, \{ j_n \} \big )}, \nonumber
\ees
where
\besn
K_\Lambda ( \{a_n\}, \{b_n\}) & = &  \sum_{n \in \mathbb{Z}^d}  \frac{e^{- \frac{n^2}{\Lambda}} }{n^2 +1 } \overline{a_n} b_n  \\
K_{\Lambda, \Lambda'}( \{a_n\}, \{b_n\}) & = &  \sum_{n \in \mathbb{Z}^d} \frac{e^{- \frac{n^2}{\Lambda}}  - e^{- \frac{n^2}{\Lambda'}}   }{n^2 + 1} \overline{a_n} b_n.
\eesn
\end{prop}

Before we give the proof, note that the exponentials in the formulas for $K_\Lambda$ and $K_{\Lambda, \Lambda'}$ above implement the ultraviolet cutoffs and thus formally as $\Lambda \to \infty$ we recover the heuristic expression. Incidentally, the exact expressions chosen here are not important. Instead of $e^{-x^2}$, we could have chosen any other even, nonnegative function $\alpha(x)$ on $\mathbb{R}$, such that $\alpha(0) = 1$, $\alpha$ is monotone decreasing for $x \geq 0$ and decays sufficiently fast at $\infty$, as well as $\alpha$ is differentiable at 0 (this last condition will be needed later). Note that formally $\mu_\infty$ would be the measure that we have regularized.

\begin{proof}
It is a standard result (see e.g. Theorem 2.3.1. in \cite{bogachev} or Theorem 1.12 in \cite{daprato2}) that if we have a symmetric, positive and trace class operator $Q$ on a Hilbert space $H$ with inner product $< \cdot , \cdot >$, then there exists a unique measure $\mu_Q$ on $H$ such that 

\ben
\int e^{i < j , h>} d \mu_Q[h] = \exp \big [  - \frac{1}{2} < j, Q j>  \big ].
\een

Now note that if $j$ and $\phi$ are in $V$, then we have that

\besn
\int_{\mathbb{T}^d} j(x) \phi(x) dx = \sum_{n \in \mathbb{Z}^d} \overline{j_n} a_n = << \{ e^{-|n| }j_n \}  , \{a_n\}   >>.
\eesn

If we let $Q_\Lambda$ and $Q_{\Lambda, \Lambda'}$ stand for the following operators on $V$

\besn
Q_\Lambda( \{ j_n \}) & = & \{ \frac{e^{ |n|} e^{- \frac{n^2}{\Lambda}} }{n^2 +1 } j_n \} \\
Q_{\Lambda, \Lambda'}( \{ j_n \}) & = & \{  \frac{e^{ |n|} ( e^{- \frac{n^2}{\Lambda}}  - e^{- \frac{n^2}{\Lambda'}} )  }{n^2 + 1} j_n \}
\eesn

We see that 

\besn
K_{\Lambda}(\{j_n\}, \{j_n\}) & = & << \{e^{-|n|}  j_n\} , Q_\Lambda \{e^{-|n|}  j_n \} >> \\
K_{\Lambda, \Lambda'}(\{j_n\}, \{j_n\}) & = & << \{e^{-|n|}  j_n\} , Q_{\Lambda, \Lambda'} \{ e^{-|n|}   j_n \} >> 
\eesn

The operators $Q_\Lambda$ and $Q_{\Lambda, \Lambda'}$ are obviously symmetric and positive. Moreover

\besn
\txt{Tr} (Q_\Lambda) & = & << e_0 , Q_\Lambda e_0>> + \sum_{b \in \mathcal{B}} <<e_b^+, Q_\Lambda e_b^+>> + \sum_{b \in \mathcal{B}} <<e_b^-, Q_\Lambda e_b^->> \\
& = & \sum_{(n_1, \dots, n_d) \in \mathbb{Z}^d} \frac{e^{ |n|} e^{- \frac{n^2}{\Lambda}} }{n^2 +1 } < \infty
\eesn

Similarly $\txt{Tr}(Q_{\Lambda, \Lambda'}) < \infty$. We thus have that these operators are indeed trace class, and therefore define the measures which satisfy what we want.
\end{proof}

\section{Wick Products}

We now define Wick products and list several of their elementary properties which will be needed below. This is standard material \cite{salmhofer, dimock}. The reason we're giving it here (apart from the reader's convenience) is that one needs to be a little careful in using the usual formulas in our setting. This is because, due to the way we have defined our measures, the $a_n$'s are not independent random variables (recall that $\overline{a_{n}} = a_{-n}$ in $V$). \newline

Let us begin with the following

\begin{prop}
For any collection $n_1, \dots, n_k \in \mathbb{Z}^d$ and $t_1, \dots, t_k \in \mathbb{C}$ we have that 

\be
\label{eq:complex}
\int_V e^{i( t_1 a_{n_1} + \dots + t_k a_{n_k})} d\mu_\Lambda = e^{ - \frac{1}{2} \sum_{\alpha, \beta =1}^k t_\alpha t_\beta K^\Lambda_{n_\alpha, n_\beta}   },
\ee 

where 

\ben
K^\Lambda_{n_\alpha, n_\beta} =\frac{e^{- \frac{n_\alpha^2}{\Lambda}} }{n_\alpha^2 +1 }   \delta_{n_\alpha, - n_\beta}  = \int_V a_{n_\alpha} a_{n_\beta} d\mu_\Lambda 
\een

\end{prop}

\begin{proof}

We shall prove the special case when the exponent of the integrand is equal to $t_n a_n + t_{-n} a_{-n}$ with $n \neq 0$. The proof of the general case is identical as the integrations involving $a_n$ and $a_m$ go `in parallel' as long as $n \neq m$ nor $n \neq -m$, while the case $n =0$ is trivial since of course $a_0 = a_{-0}$. Now, assume that $t, s \in \mathbb{R}$, we then have\footnote{The $j$ chosen here is either a linear combination of $e_n^+$ and $e_{n}^-$ or of $e_{-n}^+$ and $e_{-n}^-$ depending whether $n$ or $-n$ is in $\mathcal{B}$.}, in view of (\ref{eqns:characteristic}), that

\ben
\int_V e^{i \Big (  (t+is) a_n + (t-is)a_{-n} \Big )} d \mu_\Lambda = \exp \Big (  - ( t^2 + s^2  )\frac{e^{- \frac{n^2}{\Lambda}} }{n^2 +1 }    \Big ).
\een

Both sides of this equation are entire in $t$ and $s$. The right trivially so, while the left as a consequence of Fernique's theorem. Therefore, the equation continues to hold for arbitrary complex $t$ and $s$. Letting $t_n = t + is$ and $t_{-n} = t - is$ we get that 

\besn
\int_V e^{i (t_n a_n + t_{-n} a_{-n})}  &= & \exp \Big ( - t_n t_{-n}  \frac{e^{- \frac{n^2}{\Lambda}} }{n^2 +1 }     \Big ) \\
& = & \exp \Big ( - \frac{1}{2} \sum_{n_\alpha \in \{-n,n\}, n_\beta \in \{-n , n\}} t_{n_\alpha} t_{n_\beta} K^\Lambda_{n_\alpha, n_\beta}   \Big ).
\eesn

Consequently, we have shown everything the proposition claims apart from the formula relating $K^\Lambda_{n_\alpha, n_\beta}$ to the integral of $a_{n_\alpha} a_{n_\beta}$. However, it follows at once by differentiating both sides of (\ref{eq:complex}) with respect to $t_\alpha$ and $t_\beta$ and then setting all the $t$'s to 0.
\end{proof}

Now, define the Wick-ordered exponential via

\besn
: e^{i (t_1 a_{n_1} + \dots + t_k a_{n_k})}:_\Lambda  & \equiv &  e^{i (t_1 a_{n_1} + \dots + t_k a_{n_k})} \bigg ( \int_V e^{i (t_1 a_{n_1} + \dots + t_k a_{n_k}) } d\mu_\Lambda \bigg )^{-1} \\
& = & e^{i (t_1 a_{n_1} + \dots + t_k a_{n_k})} e^{ \frac{1}{2} \sum_{\alpha, \beta =1}^k t_\alpha t_\beta K^\Lambda_{n_\alpha, n_\beta}   },
\eesn

The Wick monomial is defined  by

\ben
: a_{n_1} \dots a_{n_k}:_\Lambda = \frac{1}{i^k} \frac{\partial^k}{ \partial t_1 \dots \partial t_k   } \bigg (   : e^{i (t_1 a_{n_1} + \dots + t_k a_{n_k})}: _\Lambda   \bigg )_{t_1, \dots, t_k =0}.
\een

From these definitions it follows immediately that

\ben
:a_{n_1} \dots a_{n_k}:_\Lambda = : a_{n_{\sigma(1)}} \dots a_{n_{\sigma(k)}}:_\Lambda,
\een

where $\sigma$ is any permutation of $\{1,2, \dots, k\}$. That

\be
\label{eq:ortho}
\int_V : a_{n_1} \dots a_{n_k} :_\Lambda : a_{m_1}  \dots a_{m_l} : _\Lambda d\mu_\Lambda  =\delta_{k,l} \sum_{\sigma} K^\Lambda_{n_1 m_{\sigma(1)}}  \dots K^\Lambda_{n_k m_{\sigma(k)}},
 \ee

where the sum goes over all permutations of $\{1,2, \dots, k\}$, and that

\bes
\label{eq:wick}
: a_{n_1} \dots a_{n_k} :_\Lambda & = & \sum_{P} \prod_{ \{ i,j\} \in P} \Big   (  - K^\Lambda_{n_i, n_j}     \Big ) \prod_{l \notin \cup P} a_{n_l} \\
\label{eq:wick'}
a_{n_1} \dots a_{n_k}  & = & \sum_{P} \prod_{ \{ i,j\} \in P} \Big   (  K^\Lambda_{n_i, n_j}     \Big ) \prod_{l \notin \cup P} : a_{n_l}:_\Lambda
\ees

where $P$ is a collection, possibly empty, of disjoint pairs of indices from $\{1, \dots, k\}$.\newline

\section{Space of Actions}

We can now describe our space of effective actions. Consider first the space $L^2(\mu_{\Lambda}, \mathbb{C})$, i.e. the space of all complex valued functions on $V$ which are square integrable with respect to $\mu_\Lambda$. We have the following

\begin{prop}
Any element $S$ of $L^2(\mu_\Lambda, \mathbb{C})$ can be written as
\be
\label{eq:l2}
S = \sum_{k \geq 0} \sum_{n_1, \dots, n_k \in \mathbb{Z}^d} G_k(n_1, \dots, n_k) :a_{n_1} \dots a_{n_k}:_\Lambda,
\ee

where the sum on the right converges in $L^2(\mu_\Lambda, \mathbb{C})$. Moreover, the $G_k$'s are invariant under permutation of their arguments and satisfy

\be
\label{eq:norm}
\sum_{k \geq 0} \sum_{ n_1, \dots, n_k} k!  \frac{ e^{- \frac{n_1^2}{\Lambda} }}{n_1^2 + 1} \dots \frac{e^{- \frac{n_k^2}{\Lambda}}}{n_k^2 + 1} |G_k(n_1, \dots, n_k)|^2 .
\ee

\end{prop}

\begin{proof}
It follows from the Wiener chaos decomposition (see e.g. Theorem 2.9.1 in \cite{bogachev} or Theorem 9.7 in \cite{daprato2}) that \textit{complex} linear combinations of products of $(a_n + a_{-n})$ and $i(a_n - a_{-n})$ are dense in $L^2(\mu_\Lambda, \mathbb{C})$. This implies that complex linear combinations of $a_{n_1} \dots a_{n_k}$ are also dense, and thus, using (\ref{eq:wick'}) that complex linear combinations of Wick products are dense in $L^2(\mu_\Lambda, \mathbb{C})$.\newline

Now, the permutation group of $k$ objects acts naturally on $\mathbb{Z}^d \times \dots \times \mathbb{Z}^d$ via $(n_1, \dots, n_k) \to (n_{\sigma(1)}, \dots, n_{\sigma(k)})$. For every $(n_1, \dots, n_k) \in \mathbb{Z}^d \times \dots \times \mathbb{Z}^d$ denote by $O(n_1, \dots, n_k)$ and $I(n_1, \dots, n_k)$, the orbit of this element and its isotropy subgroup respectively. Also, denote by $|O(n_1, \dots, n_k)|$ and $|I(n_1, \dots, n_k)|$ the cardinalities of these objects. Finally, let $O_k$ stand for the subset of $\mathbb{Z}^d \times \dots \times \mathbb{Z}^d$ which contains a single element from every orbit.\newline

Since 

\be
\label{eq:complex}
\overline{:a_{n_1} \dots a_{n_k}:_\Lambda} = :a_{-n_1} \dots a_{-n_k}:_\Lambda,
\ee

it follows from (\ref{eq:ortho}) that if $(n_1, \dots, n_k)$, $(m_1, \dots, m_k)$ are two different elements in $O_k$ then $:a_{n_1} \dots a_{n_k}:_\Lambda$ and $:a_{m_1} \dots a_{m_k}:_\Lambda$ are orthogonal as elements in $L^2(\mu_\Lambda, \mathbb{C})$. From the density of the linear combinations of the Wick products it follows in the usual way (using Bessel's inequality and the completeness of $L^2$) that the collection

\ben
\bigg \{ :a_{n_1} \dots a_{n_k}:_\Lambda \bigg \}_{k \geq 0;  (n_1, \dots, n_k) \in O_k }
\een

is an orthogonal basis for $L^2(\mu_\Lambda, \mathbb{C})$, and thus, for any elements $S \in L^2(\mu_\Lambda, \mathbb{C})$ we have the following equality

\be
\label{eq:cexpansion}
S = \sum_{k \geq 0} \sum_{(n_1, \dots, n_k) \in O_k} C_k(n_1, \dots, n_k) :a_{n_1} \dots a_{n_k}:_\Lambda,
\ee

for some coefficients $C_k(n_1, \dots, n_k)$. The sum on the right hand side converges in $L^2(\mu_\Lambda, \mathbb{C})$ and additionally, we have the following equality

\ben
|| S||_{L^2} = \sum_{k \geq 0} \sum_{(n_1, \dots, n_k) \in O_k} |C_k(n_1, \dots, n_k)|^2 || :a_{n_1} \dots a_{n_k}:_\Lambda||_{L^2}.
\een

Now, define $G_k(n_1, \dots, n_k)$ to be equal to $\frac{C(n_1, \dots, n_k)}{| O(n_1, \dots, n_k) |}$ when $(n_1, \dots, n_k) \in O_k$ and extend it to all of $\mathbb{Z}^d \times \dots \times \mathbb{Z}^d$ by stipulating that it should be constant on the orbits of the permutation group. Note that the $G_k$'s are invariant under the permutation group by construction, and we have that

\besn
S & = & \sum_{k \geq 0} \sum_{(n_1, \dots, n_k) \in O_k} | O(n_1, \dots, n_k) | G_k(n_1, \dots, n_k) :a_{n_1} \dots a_{n_k}:_\Lambda \\
& = & \sum_{k \geq 0} \sum_{(n_1, \dots, n_k) \in \mathbb{Z}^d} G_k(n_1, \dots, n_k) : a_{n_1} \dots a_{n_k}:_\Lambda
\eesn

Moreover

\besn
|| S||_{L^2} & = & \sum_{k \geq 0} \sum_{(n_1, \dots, n_k) \in O_k} |C_k(n_1, \dots, n_k)|^2 || :a_{n_1} \dots a_{n_k}:_\Lambda||_{L^2} \\
& =& \sum_{k \geq 0} \sum_{(n_1, \dots, n_k) \in O_k} || O(n_1, \dots, n_k) | G_k(n_1, \dots, n_k)|^2 || :a_{n_1} \dots a_{n_k}:_\Lambda||_{L^2} \\
& = & \sum_{k \geq 0} \sum_{(n_1, \dots, n_k) \in \mathbb{Z}^d} | O(n_1, \dots, n_k)|  | G_k(n_1, \dots, n_k)|^2 || :a_{n_1} \dots a_{n_k}:_\Lambda||_{L^2}  \\
& = &  \sum_{k \geq 0} \sum_{(n_1, \dots, n_k) \in \mathbb{Z}^d} \frac{ k!  | G_k(n_1, \dots, n_k)|^2 }{  |I(n_1, \dots, n_k)| }|| :a_{n_1} \dots a_{n_k}:_\Lambda||_{L^2} \\
& = & \sum_{k \geq 0} \sum_{ n_1, \dots, n_k} k!   |G_k(n_1, \dots, n_k)|^2 \frac{ e^{- \frac{n_1^2}{\Lambda} }}{n_1^2 + 1} \dots \frac{e^{- \frac{n_k^2}{\Lambda}}}{n_k^2 + 1}, 
\eesn

where we've used (\ref{eq:ortho}) for the last equality. The proof is thus complete.
\end{proof}

\begin{corollary*}
Any element $S$ of $L^2(\mu_\Lambda, \mathbb{R})$ can be written as in (\ref{eq:l2}) where the $G_k$'s are invariant under permutation of their arguments and satisfy 

\ben
\overline{G_k(n_1, \dots, n_k)} = G_k(-n_1, \dots, -n_k).
\een

Moreover, equation (\ref{eq:norm}) still holds.

\end{corollary*}

\begin{proof}
This follows at once from the previous proposition and from (\ref{eq:complex}).
\end{proof}

Since we will be working mostly with real-valued functions in what follows, unless stated otherwise $L^2(\mu_\Lambda)$ stands for $L^2(\mu_\Lambda, \mathbb{R})$.  

\section{Quasilocality}

Note that any functional of the form

\be
\label{eq:local}
\int_{\mathbb{T}^d} \partial^{\alpha^{(1)}} \phi(x) \dots \partial^{\alpha^{(k)}} \phi(x)   dx,
\ee

where each of $\alpha^{(1)}, \dots, \alpha^{(k)}$ is a multiindex, is in fact a well-defined element of $L^2(\mu_\Lambda)$. This is a consequence of the  inequality

\be
\label{eq:estimate}
| \partial^{\alpha} \phi(x) | \lesssim \sum_{n \in \mathbb{Z}^d} (n_1^2 + \dots n_d^2)^{|\alpha|} |a_n| \lesssim |||  a_n |||,
\ee

where $\lesssim$ means less than a constant multiple of. It thus follows that the square of the expression (\ref{eq:local}) above is bounded from above by a constant multiple of $||| a_n |||^{2k}   $. As follows by Fernique's theorem, each such expression is integrable, and we have our claim. \newline

We can now make an important 

\begin{definition}
A finite linear combination of functionals of the form (\ref{eq:local}) is called a local functional. An element in the closure in $L^2(\mu_\Lambda)$ of the set of local functionals is said to be quasilocal. The set of all quasilocal functionals will be denoted by $\mathcal{Q}(\Lambda)$.
\end{definition}

The next proposition shows that we could have defined the notion of quasilocality differently.

\begin{prop}
\label{prop:quasi}
An element of $L^2(\mu_\Lambda)$ is quasilocal if and only if in (\ref{eq:l2}) we have that
\be
\label{eq:quasilocal}
n_1 + \dots + n_k \neq 0 \implies G_k(n_1, \dots, n_k ) = 0, \quad \forall \, n_1, \dots n_k \in \mathbb{Z}^d, k \in \mathbb{N}
\ee
\end{prop}

\begin{proof}
The only if direction is a consequence of the fact that every local functional satisfies (\ref{eq:quasilocal}). This is done by a direct computation using the facts that all the fields and their derivatives are multiplied at the same point in a local functional, and that the propagator between $a_n$ and $a_m$ (which enters in the Wick product) vanishes unless $n = -m$. The fact that a limit of functionals satisfying (\ref{eq:quasilocal}), itself satisfies (\ref{eq:quasilocal}), follows at once from the expression for the $L^2$ norm given in (\ref{eq:norm}). \newline

To see the other direction, first note that for a fixed $k$, if $G_k(n_1, \dots, n_k)$ is a polynomial (in $n_1, \dots, n_k$) then, a direct computation shows that $$\sum_{n_1, \dots, n_k \in \mathbb{Z}^d} G_k(n_1, \dots, n_k) :a_{n_1} \dots a_{n_k}:_\Lambda$$ is a local functional. Thus, we will be done if we show that polynomials in $n_1, \dots, n_k$ are dense in the weighted $l^2(\mathbb{Z}^d \times \dots \times \mathbb{Z}^d, \mathbb{C})$, with the weight given by the expression before $|G_k|^2$ in (\ref{eq:norm}). However, this is a well-known fact\footnote{For reader's convenience, we give a short proof in Appendix A.} about the density in $L^2$ of polynomials for measures with sub-exponential tails (see e.g. \cite{density}). Hence, the proof is complete.
\end{proof}

In the setting of this paper, there is yet another equivalent characterization of the quasilocality, which is via translation invariance. The action of $\mathbb{R}^d$ on itself by translation ($\tau_y(x) = x+y$) descends to $\mathbb{T}^d$ (considered as a quotient space of $\mathbb{R}^d$). This in turn induces an action on the space of functions $V$, via $(\tau_y (\phi))(x) = \phi(\tau_y(x))$, which finally defines the action on the space of effective actions\footnote{Note that the word ``action'' is used with two different meanings here.} by $\tau_y (S)[\phi] = S[ \tau_y(\phi)]$. We can now make the following

\begin{definition}
An element $S$ of $L^2(\mu_\Lambda)$ is translation invariant if $\tau_y(S) = S$ for all $y \in \mathbb{R}^d$.
\end{definition}

We then have

\begin{prop}

If $S \in L^2(\mu_\Lambda)$, then $S$ is quasilocal $\iff$ $S$ is translation invariant
\end{prop}

\begin{proof}
First note that in view of the Fourier correspondence we have that if $\phi(x) \longleftrightarrow \{a_n\}_{n \in \mathbb{Z}^d}$ then $\tau_y(\phi)(x) \longleftrightarrow  \{ e^{i n y}a_n \}_{n \in \mathbb{Z}^d}$. We thus obtain a representation of the action of the translation group in the Fourier space via $\tau_y( \{a_n \}_{n \in \mathbb{Z}^d}) =  \{ e^{i n y}a_n \}_{n \in \mathbb{Z}^d}$. There are two consequences of this. \newline 

First, as is immediate from (\ref{eqns:characteristic}) the measure $\mu_\Lambda$ is translation invariant, i.e. that $\int_V \tau_y F[\phi] d \mu_\Lambda[\phi] = \int_V F[ \phi] d \mu_\Lambda[\phi]$ for any $F \in L^1(\mu_\Lambda, \mathbb{C})$ and any $y \in \mathbb{R}^d$. This in turn means that if $F_n \to F$ in $L^1(\mu_\Lambda, \mathbb{C})$ then $\tau_y(F_n) \to \tau_y(F)$ in $L^1(\mu_\Lambda, \mathbb{C})$ as well. \newline

Second, we have that 

\besn
\tau_y\Big ( :a_{n_1} \dots a_{n_k}:_\Lambda   \Big ) & = & : \tau_y(a_{n_1}) \dots \tau_y(a_{n_k}):_\Lambda \\
& = & :e^{i (n_1+ \dots + n_k)y} a_{n_1} \dots a_{n_k}:_\Lambda \\
& = & e^{i (n_1+ \dots + n_k)y}  :a_{n_1} \dots a_{n_k}:_\Lambda, 
\eesn

where the last equality is justified by (\ref{eq:wick}), since any pair of indices $n_i$ and $n_j$ which is replaced by the propagator $K_{n_i, n_j}$ satisfies $n_i + n_j = 0$. It is now easy to prove the proposition. For the only if direction, note that if $S$ is quasilocal then in view of proposition (\ref{prop:quasi})

\besn
\tau_y S & = & \tau_y \bigg (   \sum_{k \geq 0} \sum_{n_1, \dots, n_k \in \mathbb{Z}^d; n_1 + \dots + n_k = 0} G_k(n_1, \dots, n_k) :a_{n_1} \dots a_{n_k}:_\Lambda  \bigg ) \\
& = &  \sum_{k \geq 0} \sum_{n_1, \dots, n_k \in \mathbb{Z}^d; n_1 + \dots + n_k = 0} G_k(n_1, \dots, n_k) \tau_y \Big (  :a_{n_1} \dots a_{n_k}:_\Lambda \Big ) \\
&  = & \sum_{k \geq 0} \sum_{n_1, \dots, n_k \in \mathbb{Z}^d; n_1 + \dots + n_k = 0} e^{i (n_1 + \dots + n_k) y} G_k(n_1, \dots, n_k) :a_{n_1} \dots a_{n_k}:_\Lambda \\
& = & \sum_{k \geq 0} \sum_{n_1, \dots, n_k \in \mathbb{Z}^d; n_1 + \dots + n_k = 0} G_k(n_1, \dots, n_k) :a_{n_1} \dots a_{n_k}:_\Lambda  = S,
\eesn

and thus $S$ is translation invariant. \newline

To prove the converse implication, assume that $S$ is translation invariant. Using (\ref{eq:cexpansion}) we have that

\besn
S & = & \sum_{k \geq 0 } \sum_{(n_1, \dots, n_k) \in O_k} C_k(n_1, \dots, n_k) :a_{n_1} \dots a_{n_k}:_\Lambda \\
& = & \sum_{k \geq 0 } \sum_{(n_1, \dots, n_k) \in O_k} e^{i(n_1  + \dots + n_k) y} C_k(n_1, \dots, n_k) :a_{n_1} \dots a_{n_k}:_\Lambda \\
& =& \tau_yS.
\eesn

Since any two terms in the sum above are orthogonal (in $L^2(\mu_\Lambda, \mathbb{C})$), we should have that

\ben
C_k(n_1, \dots, n_k) = e^{i(n_1  + \dots + n_k) y} C_k(n_1, \dots, n_k),
\een

for every $k \geq 0$, $(n_1, \dots, n_k) \in O_k$ and $y \in \mathbb{R}^d$. If $n_1 + \dots + n_k \neq 0$, then letting $y = \frac{\pi}{n^2}n$ we get that $C_k(n_1, \dots, n_k) = - C_k(n_1, \dots, n_k)$ and thus $C_k(n_1, \dots, n_k) = 0$. It thus follows that $G_k$ vanishes on the orbit of $(n_1, \dots, n_k)$ under the permutation group and the proof is complete.
\end{proof}

At this point the reader may be wondering why on earth we have gone to such trouble defining quasilocal functionals if they simply coincide with the translation invariant ones. The reason for this is that the eventual goal of the research undertaken in this paper is to apply it to the case with no infrared cutoff. In other words, to the case when the fields do not live on the torus $\mathbb{T}^d$ but on $\mathbb{R}^d$. This is necessary if one would like to study the full renormalization group procedure rigorously, as it involves a rescaling step, which is not possible to perform on the torus. If our fields live on $\mathbb{R}^d$, the quasilocal functionals are naturally\cite{rosten} defined to have the following expansion\footnote{We are not giving here the precise meaning in which the infinite sum is meant to converge as this would be the subject of future work. In the heuristic discussion below we are only concerned with a single term.} 

\ben
 \sum_{k \geq 0} \int_{\mathbb{R}^d \times \dots \times \mathbb{R}^d} \delta(p_1 + \dots + p_k) G_k(p_1, \dots, p_k)  \phi(p_1) \dots \phi(p_k) dp_1 \dots dp_k,
\een

where the $G_k$'s are \textit{analytic} (or at least \textit{smooth}) functions of $p_1, \dots, p_k$. It is easy to see that in this context translation invariance and quasilocality are genuinely distinct concepts with quasilocal functionals being a strict subset of those which are only translation invariant. For example the functional $\Big ( \int \phi^2(x) dx \Big)^2$ is certainly translation invariant, but it is immediate that $\delta(p_1 + \dots + p_4) G_4(p_1, \dots, p_4) = \delta(p_1+ p_2) \delta (p_3 + p_4)$ which has no solution even if we only ask $G_4$ to be a continuous function, much less smooth or analytic. \newline

In view of the above, the coincidence of quasilocality and translation invariance for functionals of fields on the torus is a red herring, an artefact of the discreteness of the Fourier transform in this case, and of the fact that we are working in the $L^2$ setting. Had we been working with fields on $\mathbb{R}^d$ \textit{or} with functionals whose $G_k(n_1, \dots, n_k)$ e.g. admit analytic continuation as functions of $n_1, \dots, n_k$ we would not have had this spurious coincidence. Therefore, we shall keep referring to elements of $\mathcal{Q}(\Lambda)$ as quasilocal functionals.\newline

Now, we want to introduce the integration-out operator of the renormalization group. This is meant to to act on an element of $L^2(\mu_{\Lambda})$, produce an element of $L^2(\mu_{\Lambda'})$, and be given by the following formula

\be
\label{eq:renorm}
\mathcal{I}_{\Lambda, \Lambda'} (S) [\psi]  = - \ln \bigg (  \int e^{- S[\phi + \psi]} d\mu_{\Lambda, \Lambda'}[\phi]      \bigg).
\ee

Clearly the expression above cannot be defined on all of $\mathcal{Q}(\Lambda)$ and one needs to restrict the space somewhat. Let $b \in \mathbb{R}$ and denote by $L^2_b(\mu_\Lambda)$ the subset of all elements in $L^2(\mu_\Lambda)$ which are bounded from below by $b$. A natural restriction to make (\ref{eq:renorm}) well-defined would be to consider only elements in $\mathcal{Q}(\Lambda) \cap L^2_b(\mu_\Lambda)$. However, for technical reasons\footnote{See footnote \ref{footnote:tech}.}, it turns out to be convenient to restrict the space a little further as is given in the following

\begin{definition}
Let $\mathcal{Q}_b(\Lambda)$ be the closure in $L^2(\mu_\Lambda)$ of the bounded elements in $\mathcal{Q}(\Lambda) \cap L^2_b(\mu_\Lambda)$.
\end{definition}

We show in Appendix B that $\mathcal{Q}_b$ is rich enough to include every local functional whose integrand (defined there) is bounded from below by $b$.

\section{Flow Invariance}

We can now state the first of our main results which is the following

\begin{theorem}
$\mathcal{I}_{\Lambda, \Lambda'}$ is a continuous map from $L^2_b(\Lambda)$ to $L^2_b(\Lambda')$.
\end{theorem}

\begin{proof}
The fact that (\ref{eq:renorm}) is well defined on $L^2_b(\mu_{\Lambda})$ and that the lower bound is preserved by $\mathcal{I}_{\Lambda, \Lambda'}$ is immediate. Let $\mathcal{I}_{\Lambda, \Lambda'}'(S)[\psi]$ be given by\footnote{The notation is due to the fact that $\mathcal{I}_{\Lambda, \Lambda'}'$ is a formal linearization of the $\mathcal{I}_{\Lambda, \Lambda'}$.}

\ben
\int S[\phi + \psi] d\mu_{\Lambda, \Lambda'}[\phi].
\een

By Jensen's inequality, we have that

\ben
\mathcal{I}_{\Lambda, \Lambda'}(S)[\psi] \leq \mathcal{I}'_{\Lambda, \Lambda'}(S)[\psi] .
\een

Let $f^+$ stand for the positive part of a function. We then have that

\besn
 & & \sqrt{\int \bigg (  \mathcal{I}_{\Lambda, \Lambda'}(S)[\psi] \bigg )^2 d\mu_{\Lambda'}[\psi]  } \leq   |b| + \sqrt{ \int \bigg ( \Big ( \mathcal{I}_{\Lambda, \Lambda'}(S)[\psi] \Big )^+ \bigg )^2 d\mu_{\Lambda'}[\psi]} \leq \\
& & |b| +  \sqrt{\int \bigg ( \Big ( \mathcal{I}'_{\Lambda, \Lambda'}(S)[\psi] \Big )^+ \bigg )^2 d\mu_{\Lambda'}[\psi] } \leq   |b| + \sqrt{ \int \bigg ( \mathcal{I}'_{\Lambda, \Lambda'}(S)[\psi]  \bigg )^2 d\mu_{\Lambda'}[\psi]  }.
\eesn

Since $K_{\Lambda'}(\{a_n\},\{b_n\})  + K_{\Lambda, \Lambda'} ( \{a_n \}, \{b_n \}) = K_{\Lambda}(\{a_n\}, \{b_n\})$, we have at once that for any $L^1(\mu_{\Lambda})$ function $F$ the following equality

\ben
\int F[\chi] d\mu_{\Lambda}[\chi] = \int F[ \phi + \psi] d\mu_{\Lambda, \Lambda'} [\phi ]d\mu_{\Lambda'} [\psi] 
\een

holds.\footnote{This is of course the fundamental equation behind the renormalization group.} \newline

Therefore, again by Jensen, we have the well-known inequality

\bes
 \int \bigg ( \mathcal{I}'_{\Lambda, \Lambda'}(S)[\psi]  \bigg )^2 d\mu_{\Lambda'}[\psi]  & \leq & \int \int \Big (  S[ \phi + \psi]   \Big )^2 d\mu_{\Lambda, \Lambda'}[\phi] d\mu_{\Lambda'}[\psi] \nonumber \\
 \label{eq:less}
 & = & \int S^2[\chi] d\mu_{\Lambda}[\chi].
\ees

Putting everything together, we get that

\ben
\sqrt{\int \bigg (   \mathcal{I}_{\Lambda, \Lambda'}(S)[\psi] \bigg )^2 d\mu_{\Lambda'}[\psi]}  \leq
|b| + \sqrt{ \int S^2 [\chi] d\mu_{\Lambda}[\chi]}.
\een

It follows that $\mathcal{I}_{\Lambda, \Lambda'}$ maps $L^2_b(\mu_{\Lambda})$ into $L^2_b(\mu_{\Lambda'})$. To show that it does so continuously, let $\{S_n \}_{n=1}^\infty$ be a sequence in $L^2_b(\mu_{\Lambda})$ converging to $S_0$. First, note that for any constant $c$, we have

\ben
\mathcal{I}_{\Lambda, \Lambda'}(S + c) [\psi] = \mathcal{I}_{\Lambda, \Lambda'}(S)[\psi] + c.
\een

Using this, and the equivalences

\besn
\mathcal{I}_{\Lambda, \Lambda'}(S_n) \to \mathcal{I}_{\Lambda, \Lambda'}(S_0) & \iff & \mathcal{I}_{\Lambda, \Lambda'}(S_n) -b \to \mathcal{I}_{\Lambda, \Lambda'}(S_0) - b \\
& \iff & \mathcal{I}_{\Lambda, \Lambda'}(S_n -b) \to \mathcal{I}_{\Lambda, \Lambda'}(S_0-b ), 
\eesn

we can assume that $b = 0$ in what follows, i.e. that all our functions are nonnegative. Now, using (\ref{eq:less}) we have immediately that $\mathcal{I}'_{\Lambda, \Lambda'}(S_n) \to \mathcal{I}'_{\Lambda, \Lambda'}(S_0)$ in $L_b^2(\mu_{\Lambda'})$. Also, using (\ref{eq:less}), the fact that $|e^{-x} - e^{-y}| \leq |x - y|$ for nonnegative $x$ and $y$, and Jensen yet again we have that 

\besn
\int \bigg (  e^{ - \mathcal{I}_{\Lambda, \Lambda'}(S_n)[\psi]}  - e^{ - \mathcal{I}_{\Lambda, \Lambda'}(S_0)[\psi]}     \bigg )^2 d\mu_{\Lambda'}[\psi] & \leq & \\
\int \int \bigg ( e^{- S_n[\phi + \psi]} - e^{- S_0[\phi + \psi]}    \bigg )^2 d\mu_{\Lambda, \Lambda'}[\phi] d\mu_{\Lambda'} [\psi]  & \leq & \\
\int \Big ( S_n[\chi] - S_0[\chi]     \Big )^2 d\mu_{\Lambda}[\chi], 
\eesn

from which we conclude that $e^{ - \mathcal{I}_{\Lambda, \Lambda'}(S_n)} \to e^{ - \mathcal{I}_{\Lambda, \Lambda'}(S_0)}$ in $L^2(\mu_{\Lambda'})$. Passing to subsequences, and using the fact that if a sequence of functions converges in $L^2$ then a subsequence of it converges almost everywhere, we get a subsequence $\{S_{n_l} \}_{l=1}^\infty$ such that $\mathcal{I}_{\Lambda, \Lambda'}(S_{n_l})$ and $\mathcal{I}'_{\Lambda, \Lambda'}(S_{n_l})$ converge almost everywhere to $\mathcal{I}_{\Lambda, \Lambda'}(S_0)$ and $\mathcal{I}'_{\Lambda, \Lambda'}(S_0)$ respectively. \newline

Combining this with the fact that $\mathcal{I}_{\Lambda, \Lambda'}(S_{n_l}) \leq \mathcal{I}'_{\Lambda, \Lambda'}(S_{n_l})$ and that $\mathcal{I}'_{\Lambda, \Lambda'}(S_{n_l})$ $ \to \mathcal{I}'_{\Lambda, \Lambda'}(S_0)$ in $L^2_0(\mu_{\Lambda'})$, we get, by the generalized dominated convergence theorem, that $\mathcal{I}_{\Lambda, \Lambda'}(S_{n_l}) \allowbreak \to \mathcal{I}_{\Lambda, \Lambda'}(S_0)$ in $L^2_0(\mu_{\Lambda'})$. Thus, we get a subsequence of $\{\mathcal{I}_{\Lambda, \Lambda'}(S_n) \}_{n=1}^\infty$ which converges to $\mathcal{I}_{\Lambda, \Lambda'}(S_0)$. Since we can repeat this argument for any subsequence of the original sequence, it follows\footnote{We're using here the fact that if every subsequence of a sequence in a metric space has a sub-subsequence which converges to $x_0$, then the whole sequence converges to $x_0$ as well.} that $\mathcal{I}_{\Lambda, \Lambda'}(S_n) \to \mathcal{I}_{\Lambda, \Lambda'}(S_0)$, and we have continuity. 
\end{proof}

Our second main result is the subject of

\begin{theorem}
$\mathcal{I}_{\Lambda, \Lambda'}$ preserves quasilocality, i.e. it maps $\mathcal{Q}_b(\Lambda)$ to $\mathcal{Q}_b(\Lambda')$
\end{theorem}

We shall give two proofs of this theorem. The first is much shorter and simpler and will in fact show that $\mathcal{I}_{\Lambda, \Lambda'}$ maps $\mathcal{Q}(\Lambda) \cap L^2_b(\mu_\Lambda)$ to $\mathcal{Q}(\Lambda') \cap L^2_b(\mu_\Lambda')$. The  reason for giving the second proof is threefold. First, it is the only proof which has a chance of generalizing when quasilocality and translation invariance do not coincide. Second, in addition to proving the invariance of the quasilocal subspace, it proves the invariance of many other subspaces of functionals which are impossible to distinguish by translation invariance alone (see the discussion on page 22). Finally, as mentioned in the introduction and discussed a little further on page 22, it proves an interesting flow-invariance result in its own right.

\begin{proof}[First proof]
Assume that $S[\phi] \in \mathcal{Q}(\Lambda) \cap L^2_b(\mu_\Lambda)$. Then

\besn
\tau_y \Big ( \mathcal{I}_{\Lambda, \Lambda'} (S) [\psi] \Big ) & = & \mathcal{I}_{\Lambda, \Lambda'} (S) [\tau_y (\psi)]  \\
& = & - \ln \bigg (  \int e^{- S[\phi + \tau_y(\psi)]} d\mu_{\Lambda, \Lambda'}[\phi]      \bigg) \\
& = & - \ln \bigg (  \int e^{- S[ \tau_y (\phi) + \tau_y(\psi)]} d\mu_{\Lambda, \Lambda'}[\phi]      \bigg) \\
& = & - \ln \bigg (  \int e^{- S[ \phi + \psi]} d\mu_{\Lambda, \Lambda'}[\phi]      \bigg) \\
& = &  \mathcal{I}_{\Lambda, \Lambda'} (S) [\psi] .
\eesn

Where to get to the third line we've used the translational invariance of $\mu_{\Lambda, \Lambda'}$ (proven in the same way as for $\mu_\Lambda$). We thus have that $ \mathcal{I}_{\Lambda, \Lambda'} (S) [\psi] $ is translation invariant and is thus $\mathcal{Q}(\Lambda)$. Combining this with the previous theorem we are done.
\end{proof}

\begin{proof}[Second proof]
In view of the definition of $\mathcal{Q}_b(\Lambda)$ and the above continuity result in Theorem 1, we can assume that $S$ is bounded. Now let $F$ stand for a finite subset of $\mathbb{Z}^d$ such that $(n_1, \dots, n_d) \in F \iff -(n_1, \dots, n_d) \in F$. Denote by $P_F$ the projection of $V$ onto those components corresponding to $F$ (i.e. the map which sets all the $a_n$'s to zero unless $n \in F$). For an $L^1(\mu_\Lambda)$ function $g$, denote by $g_F$ the cylindrical approximation obtained from $g$ by integrating out the components ``not in $F$''. More precisely,

\ben
g_F[\phi] = \int g[ P_F \phi + (I - P_F) \psi] d\mu_\Lambda[\psi],
\een

where $I$ stands for the identity map. As is well-known, $g_F \to g$ in $L^2(\mu_\Lambda)$ as $F \uparrow \mathbb{Z}^d$. Denote by $\mu_{\Lambda, F}$ the pushforward of $\mu_\Lambda$ by $P_F$. It follows that if $g^{(n)} \to g$ in $L^2(\mu_\Lambda)$, then $g^{(n)}_F \to g_F$ in $L^2(\mu_{\Lambda, F})$. Needless to say, $g_F$ can be considered to be a function of only finitely many variables (those in the image of $P_F$).  \newline

Applying this to $S$, we have that $S_F$ is bounded (with the lower bound being $b$), that $S_F \to S$ in $L^2(\mu_\Lambda)$, and, recalling that $S$ is quasilocal, that if

\ben
S = \sum_{n_1 + \dots + n_k =0, k =0, 1, \dots } G_k(n_1, \dots, n_k) :a_{n_1} \dots a_{n_k}:_\Lambda,
\een

then

\ben
S_F = \sum_{n_1 + \dots + n_k=0, k = 0, 1, \dots } G_k(n_1, \dots, n_k) \Big (  : a_{n_1} \dots a_{n_k} :_\Lambda \Big )_F,
\een

with the latter sum converging in $L^2(\mu_{\Lambda, F}, \mathbb{C})$. Note that if we consider $S_F$ to be a function of finitely many variables then, by virtue of it being in $L^2(\mu_{\Lambda, F})$, it should also have an expansion of the form

\ben
S_F = \sum_{n_1, \dots, n_k \in F, k = 0, 1, \dots } \tilde{G}_k(n_1, \dots, n_k) :a_{n_1} \dots a_{n_k}:_{\Lambda, F},
\een

where $: \cdot :_{\Lambda, F}$ stands for the Wick ordering with respect to the measure $\mu_{\Lambda, F}$. We claim that in this latter expansion, $\tilde{G}(n_1, \dots, n_k) = 0$ if $n_1 + \dots + n_k \neq 0$. For obvious reasons, we shall call $L^2(\mu_{\Lambda, F})$ functions which satisfy this property quasilocal as well. Since the space of quasilocal functions is clearly closed, we will be done if we show that $\Big ( : a_{n_1} \dots a_{n_k} :_\Lambda \Big )_F$ is quasilocal. Consider one term in the definition of $:a_{n_1} \dots a_{n_k}:_\Lambda$. It is a product of a subcollection of $a_{n_1}, \dots, a_{n_k}$'s  multiplied by a collection of $(K_\Lambda)_{n,m}$'s pairing the remaining indices (see formula (\ref{eq:wick})). Since $(K_\Lambda)_{n, m}$ is proportional to $\delta_{n, -m}$, it follows that indices which appear in the subcollection of $a_n$'s are obtained from $n_1, \dots, n_k$ by omitting pairs of opposite ones. Since $n_1 + \dots + n_k = 0$, this equation is still true for the subcollection. In view of the above, we will be done if we show $(a_{m_1} \dots a_{m_l})_F$ is quasilocal if $m_1 + \dots + m_l =0$. We now use the fact that

\ben
\int a_a^A a_b^B d\mu_\Lambda = 0 \qquad \txt{unless} \qquad A = B \qquad \txt{and} \qquad a = -b.
\een

It thus follows that $(a_{m_1} \dots a_{m_l})_F = 0$ unless the $a_n$'s that get integrated are present in pairs with equal and opposite indices. It follows that the sum of the indices of the $a_n$'s that are left over after integration is done, still is equal to zero. Putting it all together, we have that $\Big ( : a_{n_1} \dots a_{n_k} :_\Lambda \Big )_F$ is equal to a sum of terms proportional to $a_{m_1} \dots a_{m_l}$ with $m_1 + \dots + m_l = 0$. It is trivial to see that each such term is quasilocal and we have what we want.  Note that this implies that $S_F$ is also quasilocal in the original sense, i.e. considered as a function on the full space. In view of the discussion above, we see that it is enough to show that $\mathcal{I}_{\Lambda, \Lambda'}(S_F)$ is quasilocal with respect to $\mu_{\Lambda,  F}$. \newline

We have thus reduced the original problem to one defined on functions of finitely many variables. Let us show now that we can further assume that $S_F$ is smooth. To this end, let\footnote{This is of course the Ornstein-Uhlenbeck semigroup action, hence the notation.}

\ben
O(\tau) S_F [\phi] = \int S_F[ e^{ - \tau} \phi + \sqrt{1 - e^{- 2 \tau}} \psi ] d\mu_{\Lambda, F}[\psi] .
\een

It is easy to see that $O(\tau) S_F$ is smooth and bounded\footnote{\label{footnote:tech} It is to justify this step that we have defined $\mathcal{Q}_b(\Lambda)$ to be the closure of quasilocal functionals bounded from above.} and that $O(\tau)S_F \to S_F$ in $L^2(\mu_{\Lambda, F})$ as $\tau \to 0^+$. Moreover, since $O(\tau)$ is diagonal in the basis of Wick products, we have that $O(\tau)S_F$ preserves quasilocality. We have thus reduced the problem to showing that a $C^\infty$, cylindrical quasilocal function maps to a quasilocal one under $\mathcal{I}_{\Lambda, \Lambda'}$.    \newline

We now need to streamline our notation. First, since $F$ will be held fixed in the rest of this paper, it shall be omitted (e.g. we'll just write $S$ instead of $S_F$, $d\mu_\Lambda$ instead of $d\mu_{\Lambda, F}$ and so on). Second, let $\Lambda' (t) = e^{-t} \Lambda$.\footnote{The exact form here is unimportant. We can take $\Lambda'$ to be any smooth function of $t$ which is equal to $\Lambda$ for $t = 0$ and which has a strictly negative derivative everywhere.} We shall denote by $K_{n,m}(t)$ the propagator of the measure $\mu_{\Lambda, \Lambda'(t)}$, i.e.

\ben
K_{n, m}(t) = \delta_{n, -m} \frac{ e^{- \frac{n^2}{\Lambda}}  - e^{- \frac{n^2}{\Lambda'(t) }  } }{2 (n^2 + 1)}.
\een

Also, $S(t)$ will stand for $\mathcal{I}_{\Lambda, \Lambda'(t)}(S)$. Additionally, we will denote the operator $\mathcal{I}'_{\Lambda'(t_1), \Lambda'(t_2)}$ (note that this is the linearized map) by $U(t_2,t_1)$. Finally, the closed subspace of $C^1$ functions (topologized with the usual $|| \cdot ||_{C^1}$ norm) whose derivative is uniformly continuous will be denoted by $BUC^1$. \newline

We now have the following important

\begin{lemma*}
$S(t)$ satisfies the following equation

\ben
S(t) =  U( t, 0) S + \int_0^t U(t, \tau) \bigg ( 2 \sum_{n, m } \dot{K}_{n,m}(\tau) \frac{ \partial S(\tau)  }{\partial a_m}  \frac{\partial S(\tau)}{\partial a_n}   \bigg ) d\tau,
\een

where this equation holds as an equation in the Banach space of continuous functions $t \to f(t)$ where $t \in [0, \infty)$ and $f(t) \in BUC^1$.
\end{lemma*}

The formula above is of course the celebrated Polchinski equation \cite{polchinski} written in the ``variation of constants" form.

\begin{proof}[Proof of Lemma]
First, note that since $S$ is $C^\infty$, then $S(t)$ is $C^\infty$ for every $t$ and thus certainly $S(t) \in BUC^1$. Now, suppose $f$ is a smooth function. We claim that the map $t \to U(t,0) f$ is continuous on $[0, \infty)$ and continuously differentiable on $(0, \infty)$\footnote{In fact, it is $C^1$ on $[0, \infty)$ but we are not going to need that.}, where $U(t, 0)f = f_t$ is considered as an element of $BUC^1$. \newline

To see this, note first that if $\Delta t >0$, we have that

\besn
|| f_{t + \Delta t} - f_t ||_{C^1} & = & || U(t+ \Delta t, t) f_t - f_t ||_{C^1} \\
&  \leq & \int ||f_t(x + y ) - f_t(x) ||_{C^1} d\mu_{\Lambda'(t_1), \Lambda'(t_2)}(y).
\eesn

Now, since the upper bound and the modulus of continuity of $f_t$ are nonincreasing as functions of $t$, the latter expression goes to $0$ uniformly in $t$ as $\Delta t \to 0^+$, as can be shown by the usual elementary argument of breaking the integral into two parts, one with small $y$ where uniform continuity of $f_t$ and its derivative is used, and the rest, which goes to zero since $d\mu_{\Lambda'(t + \Delta t), \Lambda'(t)}$ converges to a delta function. Continuity of $t \to f_t$ follows at once. \newline

To see that $t \to f_t$ is continuously differentiable on $(0, \infty)$, first note that as long as $t_0>0$, one can interchange the integral and the derivative with respect to $t$ in $\frac{df}{dt}\big|_{t = t_0}$. This is a basic fact which follows from dominated convergence. Now, note that

\besn
\bigg | \bigg | \frac{f_{t_0 + \Delta t} - f_{t_0}}{\Delta t} - \frac{df}{dt} \Big |_{t= t_0}    \bigg  |\bigg|_{C^1}  \leq \\
  || f||_{C^1} \int \bigg | \frac{  d\mu_{\Lambda, \Lambda'(t_0 + \Delta t)} - d \mu_{\Lambda, \Lambda'(t_0)}   }{\Delta t} - \frac{d}{dt} \big ( d\mu_{\Lambda, \Lambda'(t)}  \big )   \Big |_{t=t_0}  \bigg |   &  = & \\ 
 || f||_{C_1} \int \bigg | \frac{d}{dt} \big ( d\mu_{\Lambda, \Lambda'(t)} \big ) \Big |_{t=c}  - \frac{d}{dt} \big ( d\mu_{\Lambda, \Lambda'(t)}   \big )  \Big |_{t=t_0}   \bigg |,
\eesn

where $c$ in the last equation is between $t_0$ and $t_0 + \Delta t$. It is trivial to see now that the last integral goes to zero as $\Delta t \to 0$. We thus have differentiability of the map $ t \to f_t$. The fact that $t \to \frac{df}{dt}$ is continuous follows essentially in the same way as the continuity of $t \to f_t$.\newline

Now, observe that $g \to - \ln (g)$ and $h\to e^{-h}$ are local diffeomorphisms which are inverses of each other from $BUC^1$ into itself, as long as $g$ is bounded away from 0. Putting everything together, we have that $t \to S(t)$ maps into $BUC^1$, is continuous on $[0, \infty)$, and is continuously differentiable on $(0, \infty)$. \newline

The rest of the proof is a standard argument in the theory of evolution equations in Banach spaces \cite{pazy, amann}. Consider the function $\tilde{S}(\tau) = U(t, \tau) S(\tau)$. By similar arguments to the one above, (when we showed that $t \to S(t)$ is continuous and continuously differentiable) we have that $\tilde{S}(\tau)$ is a continuous function into $BUC^1$ on $[0, \infty)$, and is continuously differentiable on $(0, \infty)$. Moreover, by a direct calculation, we have that

\ben
\frac{d\tilde{S}}{d\tau} = U(t, \tau)  \bigg ( 2 \sum_{n, m } \dot{K}_{n,m}(\tau) \frac{ \partial S(\tau)  }{\partial a_m}  \frac{\partial S(\tau)}{\partial a_n}   \bigg ) .
\een

Integrating the above equation between $t_1$ and $t_2$, and then taking the limits $t_1 \to 0^+, t_2 \to t^-$, we get what we want.
\end{proof}

The lemma above shows in effect that the function $t \to S(t)$ is a fixed point of the following variation of constants map 

\ben
\Phi(f)(t) = U(t, 0) S + \int_0^t U(t, \tau)  \bigg ( 2 \sum_{n, m } \dot{K}_{n,m}(\tau) \frac{ \partial f(\tau)  }{\partial a_m}  \frac{\partial f(\tau)}{\partial a_n}  \bigg ).
\een

Let us make this more precise. Again, we follow the general ideas of the theory of the evolution equations in Banach spaces.\footnote{The reader is encouraged here to go over the introductory discussion in \cite{amann} for the abstract setting behind the argument below.} Consider the metric space $X$ of all continuous functions on $[0, \delta]$ valued in the closed ball of center 0 and radius $R$ in $BUC^1$. We shall take $R =2 ||S||_{C^1}$ and will specify $\delta$ below. The metric on $X$ is given by

\ben
\rho(f_1, f_2) = \sup_{t \in [0, \delta]} || f_1(t) - f _2(t)||_{C^1}.
\een

Note that $\Phi(v)(t)$ is indeed in $BUC^1$. This is due to the regularizing effect of $U(t, \tau)$ inside the integral. In fact, it is straightforward to show\footnote{Recall that $U$ is basically a convolution with a Gaussian. Incidentally, this is where we would need the differentiability of $\alpha$ at 0 in case we decide to replace $e^{-x^2}$ with $\alpha(x)$ in the ultraviolet cutoff.} that the following inequality $
||U(t, \tau) g ||_{C^1} \lesssim \frac{|| g||_{C^0}}{\sqrt{t - \tau}}$, holds. \newline

Now, note that for any two $BUC^1$ functions $f_1$ and $ f_2$ we trivially have the inequality

\ben
\bigg | \bigg | 2 \sum_{n, m } \dot{K}_{n,m}(\tau)\bigg (  \frac{ \partial f_1(\tau)  }{\partial a_m}  \frac{\partial f_1(\tau)}{\partial a_n}  -  \frac{ \partial f_2(\tau)  }{\partial a_m}  \frac{\partial f_2(\tau)}{\partial a_n} \bigg )  \bigg | \bigg | _{C^0} \lesssim || f_1 - f_2 ||_{C^1},
\een

where the implicit constant depends on $R$ in general. Using the inequalities above, it is easy to show that 

\besn
\rho( \Phi(f_1), \Phi(f_2) ) & \lesssim & \Big (    \int_0^\delta \frac{1}{\sqrt{t - \tau}} d \tau \Big ) \rho (f_1, f_2).
\eesn

Since the square root is an integrable function, we can choose $\delta$ small enough such that the overall constant in front of $\rho(g_1, g_2)$ is less than $\frac{1}{2}$. Using the fact that $||U(t, 0) S||_{C^1} \leq || S||_{C^1}$, we have that 

\besn
\sup_{t \in [0, \delta]} || \Phi(g)(t) ||_{C^1} & \leq & \sup_{t \in [0, \delta]} || \Phi(g)(t) - \Phi(0) ||_{C^1} + \sup_{t \in [0, \delta]} || \Phi(0) ||_{C^1} \\
& \leq & \frac{1}{2} \rho(g, 0) + \sup_{t \in [0, \delta]} || U(t, 0) S||_{C^1} < R.
\eesn

Putting everything together, we have that $\Phi$ is a contraction which takes the space $X$ to itself. We thus have that $t \to S(t)$ is indeed the unique fixed point of $\Phi$. Moreover, by starting with any point in the space and repeatedly applying $\Phi$, we converge to this fixed point. We are going to take the function $t \to U(t,0)S$ as our starting point. \newline

Now, let us say that $t \to f(t)$ in $X$ is quasilocal if for every $t$, $f(t)$ is quasilocal \emph{with respect to the appropriate measure}, i.e. with respect to the measure $\mu_{\Lambda'(t)}$. Note that according to this definition, our starting point of the iteration $t \to U(t,0)S$ is quasilocal. We now claim that if $t \to f(t)$ is quasilocal then its image by $\Phi$ is quasilocal as well. In view of the above, this will immediately imply that $S(t)$ is quasilocal, for $t \in [0, \delta]$, and thus, we will have that $S(\delta)$ is quasilocal. Repeating the argument\footnote{Note that $||e^{-S(t)}||_{C^1}$ doesn't increase as a function of $t$, and thus $||S(t) ||_{C^1}$ is bounded from above by some multiple (which is a function of $t$) of $||S||_{C^1}$. This implies that the estimates that went into showing that $\Phi$ is a contraction which preserves the space $X$, can be carried through on any compact subinterval of $[0, \infty)$.  },  we can extend this fact to all of $[0, \infty)$ and obtain that $S(t)$ is quasilocal for every $t$. This would conclude the proof of the theorem. \newline

It thus remains to demonstrate the claim. Since $U(t_2, t_1)$ acts diagonally in the Wick expansion, and thus preserves quasilocality, we will be done if we show that 

\ben
 2 \sum_{n, m } \dot{K}_{n,m}(\tau)  \frac{ \partial f(\tau)  }{\partial a_m}  \frac{\partial f(\tau)}{\partial a_n} ,
\een

is quasilocal if $f(\tau)$ is quasilocal. Therefore, consider the expression

\be
\label{eq:coeff}
\sum_{n,m} \dot{K}_{n, m}(\tau) \int \overline{   : a_{n_1} \dots a_{n_k} :_{\Lambda'(\tau)}  } \bigg ( \frac{ \partial f(\tau)  }{\partial a_m}  \frac{\partial f(\tau)}{\partial a_n}  \bigg ) d\mu_{\Lambda'(\tau)},
\ee

for some indices $n_1, \dots, n_k$ such that $n_1 + \dots + n_k \neq 0$. We need to show that the expression above vanishes. Since $\tau$ will be fixed in what remains of the proof, we shall drop it, together with the index $\Lambda'(\tau)$, and will simply write $f$ for $f(\tau)$, $d\mu$ instead of $d\mu_{\Lambda'(\tau)}$, and so on. \newline

Let us assume now that $g$ is smooth (and not just in $BUC^1$). It is easy to show then that its Wick expansion can be differentiated term by term with the result converging to the appropriate derivative. \newline

 We now use the fact \cite{bogachev} that for a smooth function $g$, we have

\ben
\int \overline{  : a_{n_1} \dots a_{n_k} : } g d\mu \simeq \int \frac{\partial^k g}{ \partial a_{n_1} \dots \partial a_{n_k}} d\mu,
\een

where $\simeq$ means equality up to an irrelevant constant. Therefore, we have that (\ref{eq:coeff}) is equal to a finite sum of terms which are schematically of the form

\ben
\int \Big ( f  \Big )^{(A)} \Big ( f  \Big )^{(B)} d\mu,
\een

where the bracketed exponents stand for derivatives. It thus follows that if $f_k$ denotes the the $k$-th partial sum of the Wick expansion of $f$ then, since $f_k \to f$ (and thus $(f_k)^{(A)} \to (f)^{(A)}$, $(f_k)^{(B)} \to (f)^{(B)}$  in $L^2(\mu)$), we have that

\ben
\int \Big ( f_k  \Big )^{(A)} \Big ( f_k  \Big )^{(B)} d\mu \to \int \Big ( f  \Big )^{(A)} \Big ( f  \Big )^{(B)} d\mu.
\een

Putting it all together, we have that 

\besn
\sum_{n,m} \dot{K}_{n, m} \int \overline{   : a_{n_1} \dots a_{n_k} :  } \bigg ( \frac{ \partial f_k }{\partial a_m}  \frac{\partial f_k}{\partial a_n}  \bigg ) d\mu \to \\
\sum_{n,m} \dot{K}_{n, m} \int \overline{   : a_{n_1} \dots a_{n_k} :  } \bigg ( \frac{ \partial f }{\partial a_m}  \frac{\partial f}{\partial a_n}  \bigg ) d\mu.
\eesn

Rearranging now in the standard way \cite{wieczerkowski, salmhofer}, and using the fact that $K_{n, m}$ is proportional to $\delta_{n, -m}$ we have that the expression above is proportional to $\delta_{n_1 + \dots + n_k, 0} = 0$, since we've assumed that $n_1 + \dots + n_k \neq 0$. The proof is thus complete for the case of a smooth $f$. \newline

Now, using the fact that $O(\sigma) f \to f$ in $L^4(\mu)$ as $\sigma \to 0^+$, we have that 

\besn
\sum_{n,m} \dot{K}_{n, m} \int \overline{   : a_{n_1} \dots a_{n_k} :  } \bigg ( \frac{ \partial (O(\sigma) f) }{\partial a_m}  \frac{\partial (O(\sigma) f)}{\partial a_n}  \bigg ) d\mu \to \\
\sum_{n,m} \dot{K}_{n, m} \int \overline{   : a_{n_1} \dots a_{n_k} :  } \bigg ( \frac{ \partial f }{\partial a_m}  \frac{\partial f}{\partial a_n}  \bigg ) d\mu.
\eesn

Since the Ornstein-Uhlenbeck semigroup preserves quasilocality, and since the limit of a zero sequence is zero, the proof is complete.
\end{proof}

We finish with a couple of remarks:

\begin{itemize}
\item Note that the latter part of the second proof above can be considered to be a theorem in  viability theory. In effect, we show that a mild solution \cite{pazy, amann} of the Polchinski's equation

\ben
\frac{\partial S}{\partial t} = 2 \sum_{n, m } \dot{K}_{n,m}(\tau) \frac{\partial^2 S}{\partial a_n \partial a_m}   - \frac{ \partial S(\tau)  }{\partial a_m}  \frac{\partial S(\tau)}{\partial a_n} ,
\een

 which starts in $\mathcal{Q}_b$, remains there. It is easy to see that the evolution generated by the linear term in the PDE above preserves quasilocality (this is simply the map $U(t_1, t_2)$ above). We do have the knowledge that the nonlinear term when acting on a quasilocal $S$ would give a quasilocal expression. However, it doesn't seem easy to control this term away from the quasilocal subset which is what is usually required in the viability literature \cite{barbu, daprato}. The proof above circumvents this by using a variation of constants formula and the smoothing effect of the $U$. It is clear that the argument above does not depend on the precise form of the Polchinski's equation, but rather on the smoothing effect of the $U$, and thus generalizes to any setting where the same fact holds.
 \item Also, note that the second proof of Theorem 2 above shows that the flow preserves much more than quasilocality (or translation invariance). As an example, consider the subspace of $\mathcal{Q}_b(\Lambda)$ consisting of all those functionals whose $G_k$'s satisfy the following stronger version of the condition in proposition (\ref{prop:quasi}): 
 
\textit{If there does not exist a decomposition of $\{1, 2, \dots, k\}$ into a collection of disjoint pairs such that for any pair $\{\alpha, \beta \}$ in this decomposition we have $n_\alpha + n_\beta = 0$, then $G_k(n_1, \dots, n_k) = 0$.}

Note that this is indeed stronger than quasilocality/translation invariance as e.g. it requires that $G_4(1, 2, 3, -6) = 0$ while such a component would not vanish for a general quasilocal element ($1+2+3+(-6) =0$ and hence it can be nonzero as far as translation invariance is concerned).The second proof above carries over trivially (this essentially boils down to the fact that $K_{n,m}$ vanishes unless $n+m=0$) and shows that this subspace is also preserved by the flow. It is easy to generalize this example to other invariant subspaces by asking for a partition of $\{1, 2, \dots, k\}$ whose elements have different even cardinalities (for example one subset has 4 elements while all the rest have 2).  \end{itemize}

\section*{Appendix A}

We want to prove here the following

\begin{prop}
Suppose $\mu( n_1, \dots, n_k) $ is a real-valued positive function on $\mathbb{Z}^d \times  \dots \times \mathbb{Z}^d$ which satisfies $|\mu(n_1, \dots, n_k) | \lesssim e^{ - A |n|}$ for every $A$ (i.e. that $\mu$ has subexponential decay). Then the space of polynomials in $n_1, \dots, n_k$ is dense in $L^2(\mathbb{Z}^d \times \dots \times \mathbb{Z}^d, \mu(n_1, \dots, n_k), \mathbb{C})$, i.e. in the weighted $l^2$ space with the weight given by $\mu(n_1, \dots, n_k)$.
\end{prop}

\begin{proof}
It is enough to prove that if $f \in L^2(\mathbb{Z}^d \times \dots \times \mathbb{Z}^d, \mu(n_1, \dots, n_k), \mathbb{C})$ is orthogonal to every polynomial then it is equal to 0. Therefore, assume that $f$ is such a function. Let $g(z_1, \dots, z_k)$ be given by 

\ben
g(z_1, \dots, z_k) = \sum_{(n_1, \dots, n_k) \in \mathbb{Z}^d \times \dots \times \mathbb{Z}^d} \mu(n_1, \dots, n_k) f(n_1, \dots, n_k) e^{i(n_1 z_1 + \dots + n_k z_k)}.
\een

By Cauchy-Schwarz and using the fast decay of $\mu$, $g$ is holomorphic everywhere. The fact that $f$ is orthogonal to every polynomial implies that all derivatives of $g$ vanish at the origin and hence, $g$ is equal to zero identically. Since this means that the Fourier transform of $g$ is identically zero, it follows that $\mu(n_1, \dots, n_k) f(n_1, \dots, n_k)$ is identicaly zero. Since $\mu$ is positive we have that $f =0$, and the proof is complete.

\end{proof}

\section*{Appendix B}

Recall that a local functional is a linear combination of expressions of the form (\ref{eq:local}). Clearly, any such expression is of the form

\ben
\int_{\mathbb{T}^d} \mathcal{L}\Big(\phi(x), \partial_{x_1} \phi (x), \dots, \partial^M_{x_d, x_d, \dots, x_d} \phi(x) \Big ) dx,
\een

where $\mathcal{L} (z_1, \dots, z_N)$ is some polynomial in $N$ variables.\footnote{$N = \sum_{m=0}^M \binom{m+ d - 1}{m}$.} Let us call this polynomial the \emph{integrand} of the local functional. In this appendix, we prove the following

\begin{prop}
If $S$ is a local functional whose integrand is bounded from below by $b$, then $S \in \mathcal{Q}_b(\Lambda)$.
\end{prop}

Note that this would imply that e.g. $\int_{\mathbb{T}^d}  (\partial \phi )^2 + P(\phi)$ is in $\mathcal{Q}_b(\Lambda)$ as long as $P(z)$ is a polynomial in $z$ bounded from below by $b$.

\begin{proof}

Clearly, it is enough to prove this for the case $b = 0$, which is what we are going to assume below. Let $\epsilon > 0$. Let $\beta$ be a constant (which we'll assume is greater than 1) such that 

\ben
| \partial^m_{x_{i_1}, \dots, x_{i_m}} \phi(x)| \leq \beta|||  a_n |||
\een

for all $i_1, \dots, i_m \in \{ 1, \dots, d\}$ and $m \leq M$. Using this estimate, we have that

\ben
\bigg |   \mathcal{L}\Big(\phi(x), \partial_{x_1} \phi (x), \dots, \partial^M_{x_d, x_d, \dots, x_d} \phi(x) \Big )   \bigg | \leq C_1 + C_2 |||a_n |||^A
\een

for some constants $A, C_1$ and $C_2$. Choose $R$ such that 

\ben
\sqrt{ \int_{B^c(0, R)} \Big ( C_1 + C_2 |||a_n|||^A   \Big )^2 d\mu_\Lambda      }   < \frac{\epsilon}{4},
\een

and

\ben
 (C_1 + C_2 (\beta(R+1))^A)  \mu_\Lambda( B^c(0, R)) < \frac{\epsilon}{4},
\een

where $B^c(0, R)$ is the complement of the ball of radius $R$ (with respect to the norm $||| \cdot |||$). The existence of such an $R$ follows immediately from Fernique's theorem. Now, let $T(z_1, \dots, z_N)$ be a trigonometric polynomial such that 

\begin{itemize}
\item $T$ is nonnegative.
\item $|T| \leq C_1 + C_2 (\beta(R+1))^A$.
\item $\sup_{(z_1, \dots, z_N) \in [-\beta R, \beta R]^N} | T(z_1, \dots, z_N) - \mathcal{L}(z_1, \dots, z_n)| < \frac{\epsilon}{2}$.
\end{itemize}

To see that such a $T$ exists, first restrict $ \mathcal{L}\Big(\phi(x), \partial_{x_1} \phi (x), \dots, \partial^M_{x_d, x_d, \dots, x_d} \phi(x) \Big ) $ to $[-\beta(R+1), \beta(R+1)]^N$, and then extend it periodically.\footnote{Note that the function $\mathcal{L}$ is periodized on a bigger region than the one on which it is uniformly approximated by $T$. This is done to ensure the continuity of the approximated function as all the discontinuities (if any) will appear on the boundary of the bigger region.} If we now take the Ces\` aro sum of a sufficiently far truncation of the Fourier series of this periodization, we get what we want directly from the properties of the Fej\' er kernel. Putting everything together, we have that 

\besn
\sqrt {  \int \bigg  |   \mathcal{L}\Big(\phi(x), \dots, \partial^M_{x_d, x_d, \dots, x_d} \phi(x) \Big )     -     T\Big(\phi(x), \dots, \partial^M_{x_d, x_d, \dots, x_d} \phi(x) \Big )  \bigg  |^2 d\mu_{\Lambda}    } & \leq & \\
\sqrt{ \int_{B(0,R)} \dots } + \sqrt{\int_{  B^c(0,R)} \dots }  <  
\sqrt{ \int_{B(0,R)} \dots }   & + & \\
  \sqrt{ \int_{B^c(0,R)} \Big ( C_1 + C_2 |||a_n|||^A   \Big )^2 d\mu_\Lambda      } + (C_1 + C_2 (\beta(R+1))^A)  \mu_\Lambda( B^c(0, R)) & < & \\
 \frac{\epsilon}{2} \sqrt{\mu_\Lambda(B(0, r))} + \frac{\epsilon}{2} < \epsilon .
\eesn

We thus have that $S$ can be approximated in $L^2(\mu_\Lambda)$ by elements of the form $\int _{\mathbb{T}^d} T\Big(\phi(x), \partial_{x_1} \phi (x), \dots, \partial^M_{x_d, x_d, \dots, x_d} \phi(x) \Big )  dx$. Since clearly each such element is bounded and nonnegative, we will have what we want provided we show that each such element is in $\mathcal{Q}$. This, in turn, would follow if we show that each element of the form

\be
\label{eq:trig}
\int_{\mathbb{T}^d} e^{i \alpha_0 \phi(x)} e^{i \alpha_1 \partial_{x_1} \phi(x)} \dots e^{i \alpha_{d, \dots, d} \partial^M_{x_d, \dots, x_d} \phi(x)} dx 
\ee

is in $\mathcal{Q}$, where $\alpha_0, \alpha_1, \dots, \alpha_{d, d, \dots, d}$ are constants.\footnote{We are slightly abusing the terminology here, as quasilocality was defined only for real-valued functions. However, either we trivially extend the definition to complex-valued ones, or we replace the product of exponentials by a product of sines and cosines. The estimates below would still go through.} To this end, note that

\besn
\int_{\mathbb{T}^d}  \prod_{n=0}^N \Bigg ( \prod_{i_1, \dots, i_n = 1, \dots, d} \bigg ( \sum_{l=0   }^L   \frac{\Big (i \alpha_{i_1, \dots, i_n} \partial^l_{x_{i_1}, \dots, x_{i_n}} \phi(x) \Big )^l   }{l!}   \bigg )   \Bigg ) dx
\eesn

is in $\mathcal{Q}$ for every $L$, converges pointwise to (\ref{eq:trig}), and is bounded (uniformly in $L$) from above by 

\ben
e^{  \big ( |\alpha_0| + |\alpha_1| + \dots + |\alpha_{d, \dots, d}|   \big ) |||a_n|||  }.
\een

By another application of Fernique's theorem, we have that they converge to (\ref{eq:trig}) in $L^2(\mu_\Lambda)$, and the proof is complete.
\end{proof}

\textbf{Acknowledgments:} The author would like to thank J. Merhej for reading a preliminary version of this paper and for the numerous comments which greatly improved its readability, as well as an anonymous referee for a detailed and thorough review.

\texttt{{\footnotesize Department of Mathematics, American University of Beirut, Beirut, Lebanon.}
}\\ \texttt{\footnotesize{Email address}} : \textbf{\footnotesize{tamer.tlas@aub.edu.lb}}


\begin{thebibliography}{99}
\bibitem{bk} D. C Brydges, T. Kennedy, ``Mayer expansions and the Hamilton-Jacobi equation'', Journal of Statistical Physics, 48, (1987), 19--49.
\bibitem{costello} K. Costello, ``Renormalization and effective field theory'', AMS, (2011).
\bibitem{rosten} O. Rosten, ``Fundamentals of the exact renormalization group'', Phys. Rep. 511, (2012), no. 4, 177--272.
\bibitem{pavel} D. Motreanu, N. Pavel, ``Tangency, flow invariance for differential equations and optimization problems'', Marcel Dekker, Inc., New York, (1999).
\bibitem{vrabie} O. C\^arj\u a, M. Necula, I. Vrabie, ``Viability, invariance and applications'', Elsevier Science B. V., Amsterdam, (2007).
\bibitem{barbu} V. Barbu, N. Pavel, ``Flow-invariant closed sets with respect to nonlinear semigroup flows'', Nonlinear Differential Equations Appl. 10, (2003), no.1, 57--72.
\bibitem{daprato} P. Cannarsa, G. Da Prato, H. Frankowska, ``Invariance for quasi-dissipative systems in Banach spaces'', J. Math. Anal. Appl. 457, (2018), no. 2, 1173--1187.
\bibitem{bogachev} V. Bogachev, ``Gaussian measures'', AMS, Providence, RI, (1998).
\bibitem{daprato2} G. Da Prato, ``An introduction to infinite-dimensional analysis'', Springer-Verlag, Berlin, (2006).
\bibitem{salmhofer} M. Salmhofer, ``Renormalization. An introduction', Texts and Monographs in Physics, Springer-Verlag, Berlin, (1999).
\bibitem{dimock} J. Dimock, ``Quantum mechanics and quantum field theory. A mathematical primer'', Cambridge University Press, Cambridge, (2011).
\bibitem{density} R. Dobru\v sin, R. Minlos, ``Polynomials of linear random functions'', Uspehi Mat. Nauk, 32, (1977), no.2 (194), 67--122, 263.
\bibitem{polchinski} J. Polchinski, ``Renormalization and Effective Lagrangians'', Nucl. Phys. B 231, (1984), 269--295.
\bibitem{pazy} A. Pazy, ``Semigroups of linear operators and applications to partial differential equations", Applied Mathematical Sciences, 44, Springer Verlag, New York, (1983).
\bibitem{amann} H. Amann, ``Linear and quasilinear parabolic problems. Vol. I'', Monographs in Mathematics, 89, Birkh\"auser, (1995).
\bibitem{wieczerkowski} C. Wieczerkowski, ``Symanzik's improved actions from the viewpoints of the renormalization group'', Comm. Math. Phys. 120, (1988), no. 1, 149--176.
\end{thebibliography}
\end{document}